\documentclass{ws-ijmpb}

\begin{document}

\markboth{SHIRO KAWABATA}
{NUMERICAL STUDY OF $\pi$-JUNCTION USING SPIN FILTERING BARRIERS}

\catchline{}{}{}{}{}
%
%%%%%%%%%%%%%%%%%%%%%%%%%%%%%%%%%%%%%%%%%%%%%%%%%%%%%%%%%%%%%%%%%%%%

\title{NUMERICAL STUDY OF $\pi$-JUNCTION USING SPIN FILTERING BARRIERS}

\author{SHIRO KAWABATA}

\address{Nanotechnology Research Institute (NRI), National Institute of Advanced Industrial Science and Technology (AIST), Tsukuba, Ibaraki, 305-8568, Japan, and \\ CREST, Japan Science and Technology Corporation (JST), Kawaguchi, Saitama 332-0012, Japan\\
s-kawabata@aist.go.jp}

\author{YASUHIRO ASANO}

\address{Department of Applied Physics, Hokkaido University,
Sapporo, 060-8628, Japan
\\
asano@eng.hokudai.ac.jp}

%\author{SECOND AUTHOR}

%\address{Group, Laboratory, Address\\
%City, State ZIP/Zone, Country\\
%secondauthor\_id@domain\_name}

\maketitle

\begin{history}
%\received{29 Oct. 2008}
%\revised{Day Month 2008}
%\accepted{(Day Month Year)}
%\comby{(xxxxxxxxxx)}
\end{history}

\begin{abstract}
We numerically investigate the Josephson transport through ferromagnetic insulators  (FIs) by taking into account its band structure.
By use of the recursive Green's function method, we found the formation of the $\pi$ junction in the case of the fully spin-polarized FI (FPFI), e.g., La${}_2$BaCuO${}_5$.
Moreover, the 0-$\pi$ transition is induced  by increasing the thickness of FPFI.
On the other hand, Josephson current through the Eu chalcogenides shows the $\pi$ junction behavior in the case of  the strong $d$-$f$ hybridization between the conduction $d$  and the localized $f$ electrons of Eu. 
Such FI-based Josephson  junctions may become a  element in the architecture of future quantum information devices.
\end{abstract}

\keywords{Josephson effects; $\pi$ junction; spin filter; spintronics; Green's function method; quantum computer.}

\section{Introduction}
Ferromagnet-superconductor hybrid structures exhibit novel  phenomena which have been studied extensively in the recent years\cite{rf:Golubov,rf:Buzdin1}
These systems provide the possibility for a controlled study of coexistence and competition of the ferromagnetism and the superconductivity. 
One of the most interesting effects is the possibility of forming the $\pi$ Josephson junction in superconductor/ferromagnetic-metal/superconductor (S-FM-S) heterostructures.\cite{rf:Bulaevskii,rf:Buzdin2}
In a $\pi$ junction the ground-state phase difference between two coupled superconductors is $\pi$ instead of 0 as in the ordinary 0 junctions.
 The existence of the $\pi$ junction in S-FM-S systems has been confirmed in experiment by Ryanzanov et al.\cite{rf:Ryanzanov} and by Kontos et al.\cite{rf:Kontos} 
In terms of the Josephson relationship $I_J= I_C \sin \phi$, where $\phi$ is the phase difference between the two superconductor layers, a transition from the 0 to $\pi$ states
implies a change in sign of $I_C$ from positive to negative. 
Physically, such a change in sign of $I_C$ is a consequence of a phase change in the pairing wave-function induced in the FM layer due to the proximity effect.

Recently, $quiet$ qubits consisting of a superconducting loop with a S-FM-S $\pi$ junction have been theoretically proposed.\cite{rf:Ioffe,rf:Blatter,rf:Yamashita}
In the quiet qubits, a quantum two level system (qubits) is spontaneously generated and therefore it is expected to be robust to the decoherence by the fluctuation of the external magnetic field.
From the viewpoint of the quantum dissipation, however, the structure of S-FM-S junctions is inherently identical with S-N-S junctions (N is a normal nonmagnetic metal).
Therefore a gapless quasiparticle excitation in the FM layer is inevitable.
This feature gives a strong Ohmic dissipation\cite{rf:Zaikin,rf:Schon} and the coherence time of S-FM-S quiet qubits is bound to be very short.

On the other hand, as was predicted by Tanaka and Kashiwaya,\cite{rf:Tanaka} the $\pi$ junction can be formed in Josephson junctions with ferromagnetic insulators (FIs).
By using the functional integral method,\cite{rf:Schon,rf:KawabataMQT1,rf:KawabataMQT2,rf:KawabataMQT3,rf:Yokoyama} we have theoretically proposed a superconductor phase\cite{rf:Kawabata1} and flux type qubits\cite{rf:Kawabata2,rf:Kawabata3,rf:Kawabata4} based on S-FI-S $\pi$ junctions.
Moreover we have showed that the effect of the dissipation due to the quasi-particle excitations on macroscopic quantum tunneling is negligibly small.\cite{rf:Kawabata3}
However, in  above studies, we have used  a very simple $\delta$-function model as the FI barrier.
Therefore, the correspondence between this toy mode and the actual band structure of FIs is unclear.
In this paper, we will formulate a numerical calculation method for the Josephson current through FIs by taking into account the band structure of FIs.
Then we will discuss the possibility of the formation of the $\pi$-coupling for the Josephson junction the two types of the FI, i.e., the fully polarized FI (FPFI) and the Eu chalcogenides (e.g., EuO and EuS).\cite{rf:EuO}

\section{Energy band structure of ferromagnetic insulators}

The typical density of states of FPFI for each spin direction is shown in Fig. 1(a).
One of the representative material of FPFI is undoped La${}_2$BaCuO${}_5$.\cite{rf:Mizuno,rf:Masuda}
The exchange splitting $V_\mathrm{ex}$ is estimated  to be 0.34 eV by the first-principle band calculation using the spin-polarized local density approximation.\cite{rf:LBCO}
Since the exchange splitting is larger and the bands were originally half-filled, the system becomes an FI.

 On the other hand, recently spin filtering effect are intensively studied by use of the Eu chalcogenides.\cite{rf:EuO,rf:Nagahama,rf:Santos}
 The schematic energy-band structure of  the Eu chalcogenides is shown in Fig. 1(b).
The Eu chalcogenides stand out among the FIs as ideal Heisenberg ferromagnets, with a high magnetic moment and a large exchange splitting of the conduction band for Eu $5d$-electrons.
Utilizing the exchange splitting ($V_\mathrm{ex}^d$) to filter spins, these materials produce a near-fully spin-polarized current when used as a tunnel barrier.
Of the Eu chalcogenides, EuO has the largest $V_\mathrm{ex}^d$ and the highest Curie temperature ($T_\mathrm{Curie} \sim 69$ K for bulk).

In EuO, the large saturation magnetic moment $7 \mu_B$ per Eu${}^{2+}$ originates from the seven unpaired electrons localized at the $4f$ band below the Fermi energy. 
Ferromagnetic order of the $4f$ spins causes exchange splitting of the conduction 5$d$ band, lowering (raising) the spin-up (-down) band symmetrically by $V_\mathrm{ex}^d/2$. 
%Thus, free carriers in the conduction band are spin-polarized. 
A large exchange splitting of 0.54 eV was determined by measuring the redshift of the absorption edge in single crystals of EuO cooled below $T_\mathrm{Curie}$.\cite{rf:Busch}

\begin{figure}[t]
\begin{center}
\includegraphics[width=10cm]{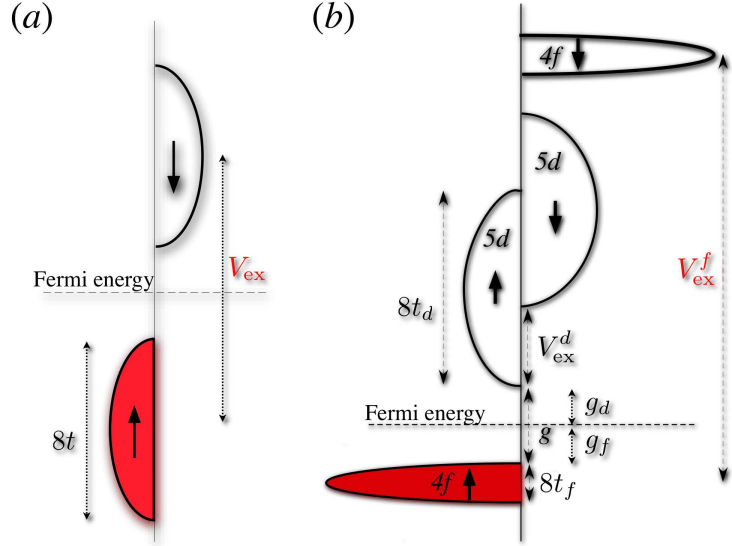}
\end{center}
\caption{The density of states for each spin direction for (a) the fully polarized ferromagnetic insulator and (b) the Eu-chalcogenides.
 }
\label{fig1}
\end{figure}

When an ultrathin film of the Eu chalcogenides is used as the tunnel barrier between two metallic electrodes, the exchange splitting of the conduction band gives rise to a lower barrier height for spin-up electrons and a
higher barrier height for spin-down electrons. 
Because of the tunnel current depends exponentially on the barrier height,\cite{rf:EuO,rf:Nagahama} the tunneling probability for spin-up electrons is much greater than for spin-down electrons, leading to a highly spin-polarized current.
This phenomenon is called the spin-filter effect.

\section{Numerical calculation of Josephson current}

In this section, we develop a numerical calculation method for the Josephson current of S-FI-S junctions.
Let us consider the two-dimensional tight-binding model for a S-FI-S junction as shown in Fig.~2.
The vector $\boldsymbol{r}=j{\boldsymbol{x}}
+m{\boldsymbol{y}}$ points to a lattice site, where ${\boldsymbol{x}}$ and ${\boldsymbol{y}}$ are unit vectors in the $x$ and $y$ directions,
respectively.
In the $y$ direction, we apply the periodic boundary condition for the number of lattice sites being $W$.

Electronic states in superconductor are described by the
mean-field Hamiltonian
 \begin{eqnarray}
{\cal H}_{\text{BCS}}&=& \frac{1}{2}\sum_{\boldsymbol{r},\boldsymbol{r}^{\prime }}%
\left[ \tilde{c}_{\boldsymbol{r}}^{\dagger }\;\hat{h}_{\boldsymbol{r},%
\boldsymbol{r}^{\prime }}\;\tilde{c}_{\boldsymbol{r}^{\prime }}^{{}}-%
\overline{\tilde{c}_{\boldsymbol{r}}}\;\hat{h}_{\boldsymbol{r},\boldsymbol{r}%
^{\prime }}^{\ast }\;\overline{\tilde{c}_{\boldsymbol{r}^{\prime }}^{\dagger
}}\;\right]  
 +\frac{1}{2}\sum_{\boldsymbol{r}\in \text{S}}\left[ \tilde{c}_{%
\boldsymbol{r}}^{\dagger }\;\hat{\Delta}\;\overline{\tilde{c}_{\boldsymbol{r}%
}^{\dagger }}-\overline{\tilde{c}_{\boldsymbol{r}}}\;\hat{\Delta}^{\ast }\;%
\tilde{c}_{\boldsymbol{r}}\right] ,  \label{bcs}
\\
\hat{h}_{\boldsymbol{r},\boldsymbol{r}^{\prime }}&=& \left[ -t \delta _{|%
\boldsymbol{r}-\boldsymbol{r}^{\prime }|,1}+(-\mu
+4t)\delta _{\boldsymbol{r},\boldsymbol{r}^{\prime }}\right] \hat{\sigma}_{0}
\end{eqnarray}
with 
$\overline{\tilde{c}}_{\boldsymbol{r}}=\left( c_{\boldsymbol{r}%
,\uparrow },c_{\boldsymbol{r},\downarrow }\right) $,
 where
  $
  c_{\boldsymbol{r} ,\sigma }^{\dagger }$ ($c_{\boldsymbol{r},\sigma }^{{}}
$)
 is the creation
(annihilation) operator of an electron at $\boldsymbol{r}$ with spin $\sigma
=$ ( $\uparrow $ or $\downarrow $ ), $\overline{\tilde{c}}$ means the
transpose of $\tilde{c}$,  and $\hat{\sigma}_{0}$ is $2\times 2$ unit matrix. 
The Fermi energy $\mu$ is set to be $4 t$ for superconductors.
In superconductors, the hopping integral $t$ is considered among nearest neighbor sites and we choose $\hat{\Delta}=i\Delta \hat{\sigma}_{2}$, where $\Delta $ is the amplitude 
of the pair potential in the $s$-wave symmetry channel, and $\hat{\sigma}_{2}$ is a Pauli matrix.

\begin{figure}[t]
\begin{center}
\includegraphics[width=10cm]{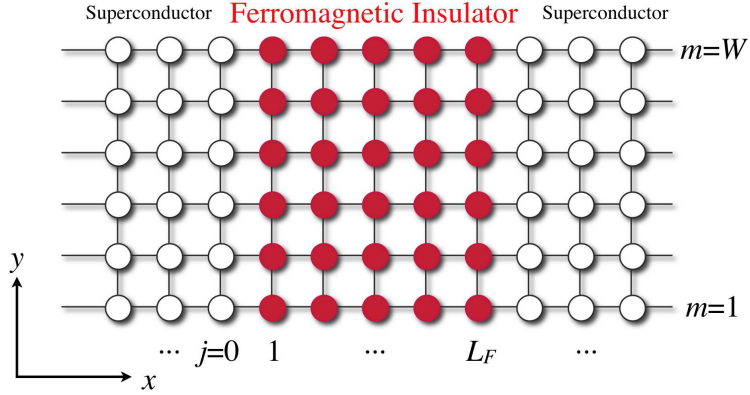}
\end{center}
\caption{A schematic figure of a Josephson junction through the ferromagnetic insulator on the
tight-binding lattice. 
 }
\label{fig2}
\end{figure}

We will consider two types of FI as a barrier of the Josephson junction.
In the case of FPFI, the Hamiltonian is given by a single-band tight-binding model as 
 \begin{eqnarray}
{\cal H}_\mathrm{FPFI} &=& -t \sum_{\boldsymbol{r},\boldsymbol{r}^{\prime },\sigma} 
c_{\boldsymbol{r},\sigma}^\dagger 
c_{\boldsymbol{r}',\sigma}
-\sum_{\boldsymbol{r}} ( 4 t -\mu)  
c_{\boldsymbol{r},\uparrow}^\dagger 
c_{\boldsymbol{r},\uparrow}
- \sum_{\boldsymbol{r}} 
( 4 t -\mu_\mathrm{FPFI} + V_\mathrm{ex}) 
 c_{\boldsymbol{r},\downarrow}^\dagger 
 c_{\boldsymbol{r},\downarrow}
,
\nonumber\\
  \end{eqnarray}
where $V_\mathrm{ex}$ is the exchange splitting [see Fig. 1(a)].
If $V_\mathrm{ex} > 4 t$, this Hamiltonian describes FPFI.
The Fermi energy $\mu_\mathrm{FPFI}$ is set to be $V_\mathrm{ex}/2  - 4t$.

On the other hand, in the case of the Eu chalcogenides, we have used a following $d$-$f$ hamiltonian,
 \begin{eqnarray}
{\cal H}_\mathrm{EC} &=& {\cal H}_d + {\cal H}_f + {\cal H}_{df}, 
\\
{\cal H}_d &=& 
-t_d \sum_{\boldsymbol{r},\boldsymbol{r}^{\prime },\sigma} 
d_{\boldsymbol{r},\sigma}^\dagger 
d_{\boldsymbol{r}',\sigma}
-\sum_{\boldsymbol{r}} ( 4 t_d -\mu_d)  
d_{\boldsymbol{r},\uparrow}^\dagger 
d_{\boldsymbol{r},\uparrow}
%\nonumber\\
%&&
- \sum_{\boldsymbol{r}} 
( 4 t_d -\mu_d + V_\mathrm{ex}^d) 
 d_{\boldsymbol{r},\downarrow}^\dagger 
 d_{\boldsymbol{r},\downarrow}
 ,
 \nonumber\\
\\
{\cal H}_f &=& 
-t_f \sum_{\boldsymbol{r},\boldsymbol{r}^{\prime },\sigma} 
f_{\boldsymbol{r},\sigma}^\dagger 
f_{\boldsymbol{r}',\sigma}
-\sum_{\boldsymbol{r}}
 ( 4 t_f -\mu_f) 
  f_{\boldsymbol{r},\uparrow}^\dagger 
  f_{\boldsymbol{r},\uparrow}
%\nonumber\\
%&&
- \sum_{\boldsymbol{r}} 
( 4 t_f -\mu_f + V_\mathrm{ex}^f)  
f_{\boldsymbol{r},\downarrow}^\dagger
 f_{\boldsymbol{r},\downarrow}
 ,
 \nonumber\\
\\
  {\cal H}_{df}
 &=& 
 V_{df} \sum_{\boldsymbol{r},\sigma} 
 \left(
 d_{\boldsymbol{r},\sigma}^\dagger  f_{\boldsymbol{r},\sigma}
 +
  f_{\boldsymbol{r},\sigma}^\dagger  d_{\boldsymbol{r},\sigma}
 \right)
 ,
\end{eqnarray}
where  $d_{\boldsymbol{r} ,\sigma }^{\dagger }$ $(f_{\boldsymbol{r} ,\sigma }^{\dagger }$) is the creation operator, $t_d$ $(t_f)$ is the hopping integral
and  $V_\mathrm{ex}^d$ $(V_\mathrm{ex}^f)$ is the exchange splitting of $d(f)$ electrons.
The Fermi energy of $d$ and $f$ electrons is respectively given by $\mu_d=-g_d$ and $\mu_f = 8 t_f + g_f$ , where $g_d$ $(g_f)$ is the energy gap of the $d(f)$ band [see Fig.1(b)].
The third term $H_{df}$ of the Hamiltonian describes the mixing between $d$ and $f$ electrons.
It was recognized for a long time that the $d$-$f$ mixing is very important to understand electronic and magnetic properties of the Eu chalcogenides.\cite{rf:Kasuya,rf:Oles,rf:Nolting1,rf:Nolting2}
So we have taken into account the $d$-$f$ mixing term in the Hamiltonian.

The Hamiltonian is diagonalized by the Bogoliubov transformation and the
Bogoliubov-de Gennes equation is numerically solved by the recursive
Green function method.\cite{rf:Furusaki,rf:Asano1,rf:Asano2} We calculate the Matsubara
Green function,
\begin{equation}
\check{G}_{\omega _{n}}(\boldsymbol{r},\boldsymbol{r}^{\prime })=\left(
\begin{array}{cc}
\hat{g}_{\omega _{n}}(\boldsymbol{r},\boldsymbol{r}^{\prime }) & \hat{f}%
_{\omega _{n}}(\boldsymbol{r},\boldsymbol{r}^{\prime }) \\
-\hat{f}_{\omega _{n}}^{\ast }(\boldsymbol{r},\boldsymbol{r}^{\prime }) & -%
\hat{g}_{\omega _{n}}^{\ast }(\boldsymbol{r},\boldsymbol{r}^{\prime })%
\end{array}
\right) , \label{deff}
\end{equation}
where $\omega _{n}=(2n+1)\pi T$ is the Matsubara frequency, $n$ is an
integer number, and $T$ is a temperature. The Josephson current is given by
\begin{equation}
I_J (\phi)=-ietT\sum_{\omega _{n}}\sum_{m=1}^{W}\mathrm{Tr}\left[ \check{G}_{\omega
_{n}}(\boldsymbol{r}^{\prime },\boldsymbol{r})-\check{G}_{\omega _{n}}(%
\boldsymbol{r},\boldsymbol{r}^{\prime })\right]
\end{equation}
with $\boldsymbol{r}^{\prime }=\boldsymbol{r}+\boldsymbol{x}$. In this
paper, $2\times 2$ and $4\times 4$ matrices are indicated by $\hat{\cdots}$
and $\check{\cdots}$, respectively. 
Throughout this paper we fix the following
parameters: $W=25$, $\mu =2t$, and $\Delta _{0}=0.01t$, $T=0.01T_{c}$ ($T_c$ is the superconductor transition temperature).

\section{Josephson current through the fully polarized ferromagnetic insulators}

\begin{figure}[t]
\begin{center}
\includegraphics[width=8.0cm]{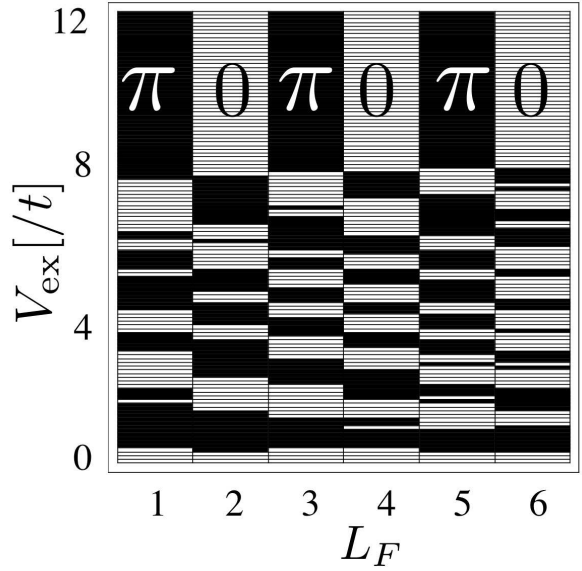}
\end{center}
\caption{The phase diagram depending on the strength of $V_\mathrm{ex}$ and $L_F$ for the FM ($0 \le V_\mathrm{ex}/t  \le 8$) and the fully polarized FI ($V_\mathrm{ex}/t > 8$). The black and white regime correspond to the $\pi$ and 0 junction, respectively.}
\label{fig3}
\end{figure}

We first discuss the Josephson current through a FPFI [Fig. 1(a)].\cite{rf:Kawabata4}
The phase diagram depending on the strength of $V_\mathrm{ex}$ ( $0 \le V_\mathrm{ex}/t  \le 8$ for FM and $V_\mathrm{ex}/t > 8$ for FI) and $L_F$ is shown in Fig.~\ref{fig3}.
The black (white) regime corresponds to the $\pi$ (0) junction [$I_J=  -(+)I_C \sin \phi$].
In the case of FPFI, the $\pi$ junction can be formed.
Moreover, the 0-$\pi$ transition is induced  by increasing the thickness of the FI barrier $L_F$.
More detailed discussion and a physical origin of above results will be given in elsewhere.\cite{rf:Kawabata6}

\section{Josephson current through the Eu-chalcogenides}

In this section, we consider the Josephson transport through the Eu-chalcogenides.\cite{rf:Kawabata6,rf:Kawabata7}
In calculation, we use the following parameters in consideration of EuO:\cite{rf:Cho,rf:Steeneken,rf:Sinjukow,rf:Ghosh,rf:Larson,rf:Kunes} $t_d=1.25$eV, $g=g_d +g_f =1.12$eV, $t_f=0.125$eV, and $V_\mathrm{ex}^d=0.528$eV.

We first discuss the Josephson current through the spin-filtering barrier only, i.e., the $d$-band [Fig. 4(a)].
The phase diagram depending on the strength of $V_\mathrm{ex}^d$ ( $0 \le V_\mathrm{ex}^d/t_d \le 6$) and the thickness of FI ($L_F$) is plotted in Fig. 4(b).
In this case, the $\pi$ junction is not formed irrespective of $L_F$ and $V_\mathrm{ex}$.
Therefore, the spin filter-effect dose not lead to the $\pi$-junction behaviors.

\begin{figure}[t]
\begin{center}
\includegraphics[width=10cm]{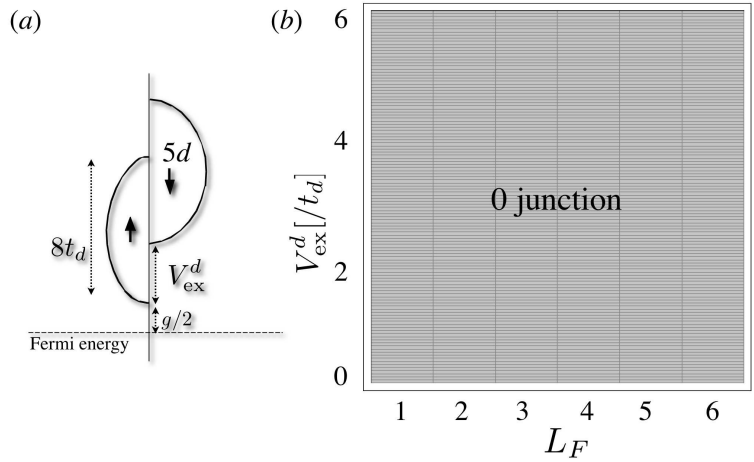}
\end{center}
\caption{(a) The density of states for each spin direction for  the spin-filtering barrier (5$d$ band of Eu).
(b) The phase diagram depending on the strength of $V_\mathrm{ex}$ and $L_F$ for the the spin-filtering barrier. 
In this case, no $\pi$ junction is formed. }
\label{fig4}
\end{figure}
\begin{figure}[bh]
\begin{center}
\includegraphics[width=8.5cm]{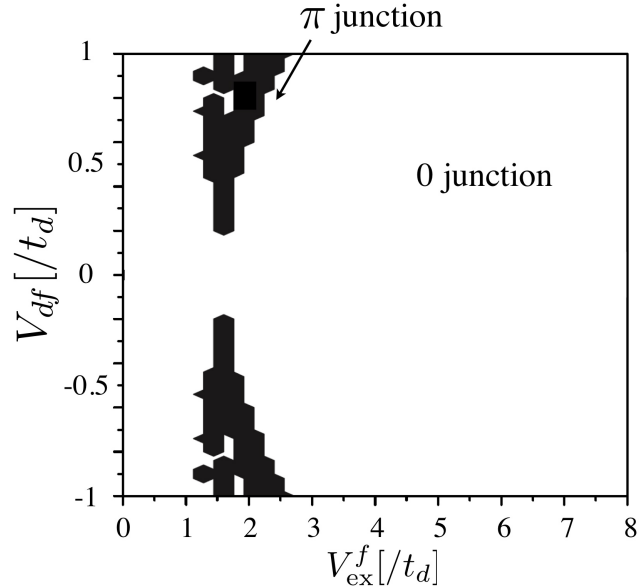}
\end{center}
\caption{The phase diagram depending on the $d$-$f$ hybridization $V_{df}$ and the exchange splitting $V_\mathrm{ex}^f$ of the $f$ band for the Josephson junction through the Eu-chalcogenides.
The black and white regime correspond to the $\pi$ and 0 junction, respectively. }
\label{fig5}
\end{figure}

Next we consider the Josephson transport through the Eu-chalcogenides with both $d$ and $f$-bands.
In calculation we set $L_F=5$ and systematically change the values of the exchange splitting of $f$ bands $V_\mathrm{ex}^f (=0.0\sim10.0$eV) and the $d$-$f$ hybridization $V_{df}(= -1.25\sim 1.25$ eV).
Fig. 5 shows the phase diagram that was numerically obtained.
The $\pi$ junction can be realized at the certain values of $V_{df}$ and $V_\mathrm{ex}^f$.
We found that the $\pi$ junction is formed if (1) $d$ and $f$ bands are overlapped each other and (2) the $d$-$f$ hybridization $V_{df}$ is strong enough.
More detailed discussion for above results will be given in elsewhere.\cite{rf:Kawabata6}

\section{Summary}
To summarize, we have studied the Josephson effect in S-FI-S junction by use of the recursive Green's function method.
We found that $\pi$ junction and the 0-$\pi$ transition is realized in the case of FPFI.
On the other hand, in the case of the Eu chalcogenides, the $\pi$ junction can be formed if the $d$ and $f$ bands are overlapped and  the $d$-$f$ hybridization is strong. 
Such FI based $\pi$ junctions may becomes a  element in the architecture of "quiet qubit".

\section*{Acknowledgements}

We  would like to thank J. Arts, A. Brinkman, M. Fogelstr\"om, A. A. Golubov, S. Kashiwaya, P. J. Kelly, T. L\"ofwander, T. Matsumoto, T. Nagahama, J. Pfeiffer, Y. Tanaka, T. Takimoto, and M. Weides for useful discussions.
This work was  supported by CREST-JST and a Grant-in-Aid for Scientific Research from the Ministry of Education, Science, Sports and Culture of Japan (Grant No. 19710085).

\end{document}